\catcode`\@=11					% To make protected \def's

%************************************************************
%*
%*		Font set-up
%*
%************************************************************

%************** 5-point fonts *******************************

\font\fiverm=cmr5				% roman
\font\fivemi=cmmi5				% math italic
\font\fivesy=cmsy5				% math symbols
\font\fivebf=cmbx5				% bold face

\skewchar\fivemi='177
\skewchar\fivesy='60

%************** 6-point fonts *******************************

\font\sixrm=cmr6				% roman
\font\sixi=cmmi6				% math italic
\font\sixsy=cmsy6				% math symbols
\font\sixbf=cmbx6				% bold face

\skewchar\sixi='177
\skewchar\sixsy='60

%************** 7-point fonts *******************************

\font\sevenrm=cmr7				% roman
\font\seveni=cmmi7				% math italic
\font\sevensy=cmsy7				% math symbols
\font\sevenit=cmti7				% italic
\font\sevenbf=cmbx7				% bold face

\skewchar\seveni='177
\skewchar\sevensy='60

%************** 8-point fonts *******************************

\font\eightrm=cmr8				% roman
\font\eighti=cmmi8				% math italic
\font\eightsy=cmsy8				% math symbols
\font\eightit=cmti8				% italic
				% slanted
\font\eightbf=cmbx8				% bold face
				% typewriter
				% sans serif

\skewchar\eighti='177
\skewchar\eightsy='60

%************** 9-point fonts *******************************

\font\ninei=cmmi9
\font\ninesy=cmsy9

\skewchar\ninei='177
\skewchar\ninesy='60

%************** 10-point fonts ******************************

\font\tenrm=cmr10				% roman
\font\teni=cmmi10				% math italic
\font\tensy=cmsy10				% math symbols

\font\tenex=cmex10				% math extension
\font\tenit=cmti10				% italic
\font\tensl=cmsl10				% slanted
\font\tenbf=cmbx10				% bold face
\font\tentt=cmtt10				% typewriter
\font\tenss=cmss10				% sans serif
\font\tensc=cmcsc10				% small caps
\font\tenbi=cmmib10				% bold math

\skewchar\teni='177
\skewchar\tenbi='177
\skewchar\tensy='60

\def\tenpoint{\ifmmode\err@badsizechange\else
	\textfont0=\tenrm \scriptfont0=\sevenrm \scriptscriptfont0=\fiverm
	\textfont1=\teni  \scriptfont1=\seveni  \scriptscriptfont1=\fivemi
	\textfont2=\tensy \scriptfont2=\sevensy \scriptscriptfont2=\fivesy
	\textfont3=\tenex \scriptfont3=\tenex   \scriptscriptfont3=\tenex
	\textfont4=\tenit \scriptfont4=\sevenit \scriptscriptfont4=\sevenit
	\textfont5=\tensl
	\textfont6=\tenbf \scriptfont6=\sevenbf \scriptscriptfont6=\fivebf
	\textfont7=\tentt
	\textfont8=\tenbi \scriptfont8=\seveni  \scriptscriptfont8=\fivemi
	\def\rm{\tenrm\fam=0 }%
	\def\it{\tenit\fam=4 }%
	\def\sl{\tensl\fam=5 }%
	\def\bf{\tenbf\fam=6 }%
	\def\tt{\tentt\fam=7 }%
	\def\ss{\tenss}%
	\def\sc{\tensc}%
	\def\bmit{\fam=8 }%
	\rm\setparameters\setbaselines\fi}

%************** 12-point fonts ******************************

\font\twelverm=cmr12				% roman
\font\twelvei=cmmi12				% math italic
\font\twelvesy=cmsy10	scaled\magstep1		% math symbols
\font\twelveex=cmex10	scaled\magstep1		% math extension
\font\twelveit=cmti12				% italic
\font\twelvesl=cmsl12				% slanted
\font\twelvebf=cmbx12				% bold face
\font\twelvett=cmtt12				% typewriter
\font\twelvess=cmss12				% sans serif
\font\twelvesc=cmcsc10	scaled\magstep1		% small caps
\font\twelvebi=cmmib10	scaled\magstep1		% bold math
	% bold italic

\skewchar\twelvei='177
\skewchar\twelvebi='177
\skewchar\twelvesy='60

\def\twelvepoint{\ifmmode\err@badsizechange\else
	\textfont0=\twelverm \scriptfont0=\eightrm \scriptscriptfont0=\sixrm
	\textfont1=\twelvei  \scriptfont1=\eighti  \scriptscriptfont1=\sixi
	\textfont2=\twelvesy \scriptfont2=\eightsy \scriptscriptfont2=\sixsy
	\textfont3=\twelveex \scriptfont3=\tenex   \scriptscriptfont3=\tenex
	\textfont4=\twelveit \scriptfont4=\eightit \scriptscriptfont4=\sevenit
	\textfont5=\twelvesl
	\textfont6=\twelvebf \scriptfont6=\eightbf \scriptscriptfont6=\sixbf
	\textfont7=\twelvett
	\textfont8=\twelvebi \scriptfont8=\eighti  \scriptscriptfont8=\sixi
	\def\rm{\twelverm\fam=0 }%
	\def\it{\twelveit\fam=4 }%
	\def\sl{\twelvesl\fam=5 }%
	\def\bf{\twelvebf\fam=6 }%
	\def\tt{\twelvett\fam=7 }%
	\def\ss{\twelvess}%
	\def\sc{\twelvesc}%
	\def\bmit{\fam=8 }%
	\rm\setparameters\setbaselines\fi}

%************** 14-point fonts ******************************

\font\fourteenrm=cmr10	scaled\magstep2		% roman -- talaris=cmr10 \step2
\font\fourteeni=cmmi10	scaled\magstep2		% math italic
\font\fourteensy=cmsy10	scaled\magstep2		% math symbols
\font\fourteenex=cmex10	scaled\magstep2		% math extension
\font\fourteenit=cmti10	scaled\magstep2		% italic
\font\fourteensl=cmsl10	scaled\magstep2		% slanted
\font\fourteenbf=cmbx10	scaled\magstep2		% bold face
\font\fourteentt=cmtt10	scaled\magstep2		% typewriter
\font\fourteenss=cmss10	scaled\magstep2		% sans serif
\font\fourteensc=cmcsc10 scaled\magstep2	% small caps
\font\fourteenbi=cmmib10 scaled\magstep2	% bold math

\skewchar\fourteeni='177
\skewchar\fourteenbi='177
\skewchar\fourteensy='60

\def\fourteenpoint{\ifmmode\err@badsizechange\else
	\textfont0=\fourteenrm \scriptfont0=\tenrm \scriptscriptfont0=\sevenrm
	\textfont1=\fourteeni  \scriptfont1=\teni  \scriptscriptfont1=\seveni
	\textfont2=\fourteensy \scriptfont2=\tensy \scriptscriptfont2=\sevensy
	\textfont3=\fourteenex \scriptfont3=\tenex \scriptscriptfont3=\tenex
	\textfont4=\fourteenit \scriptfont4=\tenit \scriptscriptfont4=\sevenit
	\textfont5=\fourteensl
	\textfont6=\fourteenbf \scriptfont6=\tenbf \scriptscriptfont6=\sevenbf
	\textfont7=\fourteentt
	\textfont8=\fourteenbi \scriptfont8=\tenbi \scriptscriptfont8=\seveni
	\def\rm{\fourteenrm\fam=0 }%
	\def\it{\fourteenit\fam=4 }%
	\def\sl{\fourteensl\fam=5 }%
	\def\bf{\fourteenbf\fam=6 }%
	\def\tt{\fourteentt\fam=7}%
	\def\ss{\fourteenss}%
	\def\sc{\fourteensc}%
	\def\bmit{\fam=8 }%
	\rm\setparameters\setbaselines\fi}

%************** Miscellaneous big fonts *********************

\font\seventeenrm=cmr10 scaled\magstep3		% roman
		% bold face

%************************************************************
%*
%*		Parameter initialization
%*
%************************************************************

\newdimen\rp@
\newcount\@basestretchnum
\newskip\@baseskip
\newskip\headskip
\newskip\footskip

% Routine to set page parameters

\def\setparameters{\rp@=.1em
	\headskip=24\rp@
	\footskip=\headskip
	\delimitershortfall=5\rp@
	\nulldelimiterspace=1.2\rp@
	\scriptspace=0.5\rp@
	\abovedisplayskip=10\rp@ plus3\rp@ minus5\rp@
	\belowdisplayskip=10\rp@ plus3\rp@ minus5\rp@
	\abovedisplayshortskip=5\rp@ plus2\rp@ minus4\rp@
	\belowdisplayshortskip=10\rp@ plus3\rp@ minus5\rp@
	\normallineskip=\rp@
	\lineskip=\normallineskip
	\normallineskiplimit=0pt
	\lineskiplimit=\normallineskiplimit
	\jot=3\rp@
	\setbox0=\hbox{\the\textfont3 B}\p@renwd=\wd0
	\skip\footins=12\rp@ plus3\rp@ minus3\rp@
	\skip\topins=0pt plus0pt minus0pt}

% Special routine to scale \baselineskip

\def\setbaselines{\maxdepth=4\rp@\baselinestretch=\@basestretchnum}

% The \baselinestretch command

\def\baselinestretch{\afterassignment\@basestretch\@basestretchnum}
\def\@basestretch{%
	\@baseskip=12\rp@ \divide\@baseskip by1000
	\normalbaselineskip=\@basestretchnum\@baseskip
	\baselineskip=\normalbaselineskip
	\bigskipamount=\the\baselineskip
		plus.25\baselineskip minus.25\baselineskip
	\medskipamount=.5\baselineskip
		plus.125\baselineskip minus.125\baselineskip
	\smallskipamount=.25\baselineskip
		plus.0625\baselineskip minus.0625\baselineskip
	\setbox\strutbox=\hbox{\vrule height.708\baselineskip
		depth.292\baselineskip width0pt }}

%************************************************************
%*
%*		Modifications to PLAIN.TEX
%*
%************************************************************

% Modifications to PLAIN routines to handle scaling of page parameters

\def\makeheadline{\vbox to0pt{\baselinestretch=1000
	\vskip-\headskip \vskip1.5pt
	\line{\vbox to\ht\strutbox{}\the\headline}\vss}\nointerlineskip}

\def\makefootline{\baselineskip=\footskip\line{\the\footline}}

\def\big#1{{\hbox{$\left#1\vbox to8.5\rp@ {}\right.\n@space$}}}
\def\Big#1{{\hbox{$\left#1\vbox to11.5\rp@ {}\right.\n@space$}}}
\def\bigg#1{{\hbox{$\left#1\vbox to14.5\rp@ {}\right.\n@space$}}}
\def\Bigg#1{{\hbox{$\left#1\vbox to17.5\rp@ {}\right.\n@space$}}}

% Modifications to PLAIN to handle bold math

\mathchardef\alpha="710B
\mathchardef\beta="710C
\mathchardef\gamma="710D
\mathchardef\delta="710E
\mathchardef\epsilon="710F
\mathchardef\zeta="7110
\mathchardef\eta="7111
\mathchardef\theta="7112
\mathchardef\iota="7113
\mathchardef\kappa="7114
\mathchardef\lambda="7115
\mathchardef\mu="7116
\mathchardef\nu="7117
\mathchardef\xi="7118
\mathchardef\pi="7119
\mathchardef\rho="711A
\mathchardef\sigma="711B
\mathchardef\tau="711C
\mathchardef\upsilon="711D
\mathchardef\phi="711E
\mathchardef\chi="711F
\mathchardef\psi="7120
\mathchardef\omega="7121
\mathchardef\varepsilon="7122
\mathchardef\vartheta="7123
\mathchardef\varpi="7124
\mathchardef\varrho="7125
\mathchardef\varsigma="7126
\mathchardef\varphi="7127
\mathchardef\imath="717B
\mathchardef\jmath="717C
\mathchardef\ell="7160
\mathchardef\wp="717D
\mathchardef\partial="7140
\mathchardef\flat="715B
\mathchardef\natural="715C
\mathchardef\sharp="715D

%************************************************************
%*
%*		Initialization
%*
%************************************************************

\def\err@badsizechange{%
	\immediate\write16{--> Size change not allowed in math mode, ignored}}

\baselinestretch=1000
\tenpoint

\catcode`\@=12					% Restore @ sign
% Routine to guarantee that this file is input only once
\catcode`\@=11
\expandafter\ifx\csname @iasmacros\endcsname\relax
	\global\let\@iasmacros=\par
\else	\endinput
\fi
\catcode`\@=12

% Set up font size commands and \baselinestretch command

% Some alternative font names
\def\rmb{\seventeenrm}

\def\itb{icmsy8}

% Simple spacing commands
\def\singlespace{\baselineskip=\normalbaselineskip}
\def\halfspace{\baselineskip=1.5\normalbaselineskip}
\def\doublespace{\baselineskip=2\normalbaselineskip}

% Macros for references and abstracts

\def\AB{\bigskip\parindent=40pt
        \centerline{\bf ABSTRACT}\medskip\halfspace\narrower}
\def\AE{\bigskip\nonarrower\doublespace}
\def\nonarrower{\advance\leftskip by-\parindent
	\advance\rightskip by-\parindent}

% Useful commands

\def\boxit#1{\vbox{\hrule\hbox{\vrule\kern3pt
	\vbox{\kern3pt#1\kern3pt}\kern3pt\vrule}\hrule}}

% Special symbols
\def\hence{\leavevmode\hbox{\bf .\raise5.5pt\hbox{.}.} }

\def\dalemb#1#2{{\vbox{\hrule height.#2pt
	\hbox{\vrule width.#2pt height#1pt \kern#1pt \vrule width.#2pt}
	\hrule height.#2pt}}}
\def\gtorder{\mathrel{\raise.3ex\hbox{$>$}\mkern-14mu
             \lower0.6ex\hbox{$\sim$}}}
\def\ltorder{\mathrel{\raise.3ex\hbox{$<$}\mkern-14mu
             \lower0.6ex\hbox{$\sim$}}}

% For twoup output
\newdimen\fullhsize
\newbox\leftcolumn
\def\twoup{\hoffset=-.5in \voffset=-.25in
  \hsize=4.75in \fullhsize=10in \vsize=6.9in
  \def\fullline{\hbox to\fullhsize}
  \let\lr=L
  \output={\if L\lr
        \global\setbox\leftcolumn=\columnbox\global\let\lr=R \advancepageno
      \else \doubleformat \global\let\lr=L\fi
    \ifnum\outputpenalty>-20000 \else\dosupereject\fi}
  \def\doubleformat{\shipout\vbox{
    \fullline{\box\leftcolumn\hfil\columnbox}\advancepageno}}
  \def\columnbox{\leftline{\vbox{\makeheadline\pagebody\makefootline}}}
  \tolerance=1000 }

\twelvepoint
\doublespace
\font\itb=cmr10 scaled 1440
{\nopagenumbers{
\rightline{IASSNS-HEP-96/02}
\rightline{~~~January, 1996}
\bigskip\bigskip
\centerline{\rmb Projective Group Representations}
\centerline{\rmb in Quaternionic Hilbert Space}
\medskip
\centerline{\itb Stephen L. Adler
%{\singlespace { Research supported in part by the Department of Ene
%rgy 
%under Grant No.
%DE-FG02-90ER40542.}}
}
\centerline{\bf Institute for Advanced Study}
\centerline{\bf Princeton, NJ 08540}
\medskip
%\leftline{{\it Short title:} short title}
\bigskip\bigskip
\leftline{\it Send correspondence to:}
\medskip
{\singlespace\leftline{Stephen L. Adler}
\leftline{Institute for Advanced Study}
\leftline{Olden Lane, Princeton, NJ 08540}
\leftline{Phone 609-734-8051; FAX 609-924-8399; email adler@sns.ias.edu}}
\bigskip\bigskip
}}
\vfill\eject
\pageno=2
\AB
We extend the discussion of projective group representations in 
qua-\hfill \break
ternionic Hilbert space which was given in our recent book.  
The associativity condition for
quaternionic projective representations is formulated in terms of unitary 
operators and then analyzed in terms
of their generator structure.  The multi--centrality and 
centrality assumptions 
are also analyzed in generator terms, and implications of this 
analysis are discussed.  
\AE
\bigskip\bigskip
\vfill\eject
\pageno=3
\leftline{\bf I. ASSOCIATIVITY CONDITION FOR QUATERNIONIC PROJECTIVE} 
\leftline{\bf ~~~GROUP REPRESENTATIONS}

In quaternionic quantum mechanics,                                                   
all symmetries of the transition probabilities
are generated by unitary transformations acting on the states of Hilbert 
space.$^{1-3}$  In the simplest
case, the unitary transformations $U_a,~U_b,...$ form a representation 
(or vector representation) of 
the symmetry group with elements $a,~b,...$,
$$U_bU_a=U_{ba}~~~.  \eqno(1)$$
A more general possibility is that the group multiplication table is 
represented over the {\it rays} corresponding to a complete set of physical 
states, but not over individual state vectors chosen as ray representatives.  
This more general composition rule defines a 
quaternionic {\it projective representation} (or ray 
representation), and takes the form (Adler$^4$, Sec. 4.3)
$$U_bU_a |f\rangle=U_{ba} |f\rangle \omega(f;b,a),~~~~~|\omega(f;b,a)|=1~~~,
\eqno(2)$$
for one particular complete set 
of states $ |f\rangle $ and a set of quaternionic 
phases $ \omega(f;b,a) $.  
When we change ray representative from $|f\rangle$ to 
$|f_{\phi}\rangle \equiv |f\rangle \phi$, with $|\phi|=1$, 
the phase defining the projective
representation is easily seen to transform as
$$\omega(f_{\phi};b,a)=\bar \phi \omega(f;b,a) \phi~~~,\eqno(3)$$
with the bar denoting quaternion conjugation.  Equation (3) shows clearly 
that the projective phase $\omega$ must depend on the state label $f$ as 
well as on the group elements $a,b$; failure to take this into account can  
lead$^4$ to erroneous conclusions (as in $^5$) concerning quaternionic 
projective representations. 

The defining relation for quaternionic projective representations given in
Eq.~(2) can be rewritten in operator form by defining a left--acting
operator $\Omega(b,a)$, 
$$\Omega(b,a)=\sum_f |f\rangle \omega(f;b,a) \langle f|~~~, \eqno(4a)$$    
which using Eq.~(3) is seen to be independent of the
ray representative chosen for the states $ |f\rangle $.  Multiplying 
Eq.~(2) from the right by $\langle f| $ and summing over the complete 
set of states $ |f\rangle $, we obtain the operator form of the 
projective representation,
$$U_bU_a=U_{ba} \Omega(b,a)~~~.\eqno(4b)$$  
It is also immediate from the definition
of Eq.~(4a), and the fact that $|\omega|=1$, that the operator $\Omega(b,a)$
is quaternion unitary,
$$\Omega(b,a)^{\dagger} \Omega(b,a) = \Omega(b,a) \Omega(b,a)^{\dagger}=1
~~~.\eqno(5)$$
Note that if we were to make the definition of a quaternionic projective 
representation more restrictive by requiring that Eq.~(2) hold for {\it all} 
states in Hilbert space, rather than for one particular complete set of 
states, then we would require $\Omega(b,a)=1$, since the unit operator is the 
only unitary operator which is simultaneously diagonal on all complete bases 
in quaternionic Hilbert space.  Hence this more restrictive definition 
excludes quaternionic embeddings of complex projective representations, 
whereas these are admitted as quaternionic projective representations by the 
definition of Eq.~(2).  

A nontrivial condition on the projective representation structure is obtained 
from the associativity of multiplication in quaternionic Hilbert space, which 
implies
$$(U_c U_b) U_a =U_c (U_b U_a)~~~.\eqno(6)$$ 
Applying Eq.~(4b) twice to the left hand side of Eq.~(6), we get 
$$\eqalign{
(U_c U_b) U_a=&U_{cb} \Omega(c,b) U_a =
U_{cb}U_a~ U_a^{-1} \Omega(c,b) U_a  \cr
=&U_{cba} \Omega(cb,a) U_a^{-1} \Omega(c,b) U_a~~~, \cr
}\eqno(7a)$$
while applying Eq.~(4b) twice to the right hand side of Eq.~(6) gives
$$\eqalign{
U_c(U_b U_a)=&U_c U_{ba} \Omega(b,a) \cr
=&U_{cba} \Omega(c,ba) \Omega(b,a) ~~~. \cr
}\eqno(7b)$$
Upon multiplying from the left by $U_{cba}^{-1}$, Eqs.~(7a,b) give the 
operator form of the {\it associativity condition} 
$$\Omega(c,ba) \Omega(b,a)=\Omega(cb,a) U_a^{-1} \Omega(c,b) U_a~~~.
\eqno(8)$$

We can also express the associativity condition as a condition on the 
quaternionic
phase $\omega(f;b,a)$ introduced in Eq.~(2), 
by applying the spectral  
representation of Eq.~(4a) to the operator form of the associativity 
condition given in Eq.~(8).  From Eq.~(4a) we get
$$\Omega(c,ba)=\sum_f |f\rangle \omega(f;c,ba)\langle f|~~~,\eqno(9a)$$
which when multiplied from the right by Eq.~(4a) gives
$$\Omega(c,ba) \Omega(b,a)
=\sum_f|f\rangle \omega(f;c,ba) \omega(f; b,a) \langle f|~~~.\eqno(9b)$$
Equation (4a) and the unitarity of $\Omega(cb,a)$ also imply that 
$$\Omega(cb,a)^{-1}=\sum_f |f\rangle \overline{\omega(f;cb,a)} \langle f|
~~~,\eqno(9c)$$
and so the associativity condition of Eq.~(8) can be rewritten as 
$$\eqalign{
U_a^{-1} \Omega(c,b) U_a=&\Omega(c,ba) \Omega(b,a) \Omega(cb,a)^{-1} \cr
=&\sum_f|f\rangle  \omega(f;c,ba) \omega(f;b,a) \overline{\omega(f;cb,a)}
\langle f|~~~.\cr
}\eqno(10)$$
Hence $U_a^{-1} \Omega(c,b) U_a$ is diagonal in the basis spanned by 
the states $|f \rangle$.  Taking matrix elements of Eq.~(10), and using the 
unitarity of $U_a$, the associativity condition gives the two relations 
$$\omega(f;c,ba) \omega(f;b,a) \overline{\omega(f;cb,a)}
=\sum_{f^{\prime \prime}} \overline{\langle f^{\prime \prime} | U_a |f \rangle}
\omega(f^{\prime \prime};c,b) 
\langle f^{\prime \prime} | U_a |f \rangle~~~,\eqno(11)$$
and, when $\langle f^{\prime} | f \rangle=0$, 
$$0=
\sum_{f^{\prime \prime}} \overline{\langle f^{\prime \prime} | U_a |f \rangle}
\omega(f^{\prime \prime};c,b) 
\langle f^{\prime \prime} | U_a |f^{\prime} \rangle~~~.
\eqno(12)$$

We conclude this section by comparing the quaternionic Hilbert space form of 
the associativity condition with the simpler form which is familiar from 
complex Hilbert space.$^{6,7}$  In a complex Hilbert space, 
the phase $\omega(f;b,a)$ introduced in Eq.~(2) is a complex number, and 
commutes with the phase $\phi$, also now complex,  which we 
introduced in Eq.~(3) to describe a
change of ray representative.  Hence Eq.~(3) implies, in the complex case, 
that $\omega(f;b,a)$ is independent of the ray representative chosen 
for the state $|f\rangle$, and 
it is then consistent to assume that $\omega(f;b,a)$ is independent of the 
state label $f$, so that
$$\omega(f;b,a)= \omega(b,a) ~~~~~{\rm complex~Hilbert~space}~~~.\eqno(13a)$$
Substituting Eq.~(13a) into Eq.~(4a), we now get
$$\Omega(b,a)=\sum_f |f\rangle \omega(b,a) \langle f|~~~
=\omega(b,a) \sum_f |f\rangle  \langle f|
=\omega(b,a) 1~~~, \eqno(13b)$$    
where $1$ denotes the unit operator in complex Hilbert space.   Since the 
complex phase 
$\omega(b,a)$ is a $c$--number in complex Hilbert space, on substituting 
Eq.~(13b) into Eq.~(4b) we learn that 
$$U_bU_a=U_{ba} \omega(b,a)=\omega(b,a) U_{ba}~~~,\eqno(14a)$$  
which is the standard definition of a projective representation in complex
Hilbert space.  Moreover, since Eq.~(13b) implies that
$\Omega(b,a)$ commutes with the unitary operator $U_a$, the associativity 
condition of Eqs.~(8) and (11) 
reduces to the familiar complex Hilbert space form 
$$\omega(c,ba) \omega(b,a)=\omega(cb,a) \omega(c,b) ~~~.
\eqno(14b)$$

\bigskip
\leftline{\bf II. THE ASSOCIATIVITY CONDITION IN GENERATOR FORM}

Let us now assume that the symmetry group with which we are dealing is a 
Lie group, so that in the neighborhood of the identity $e$ the unitary 
transformations $U_a,~U_b,~U_{ba},...$ can be written in terms of a set of 
anti--self--adjoint generators $\tilde G_A$ as 
$$U_a=\exp(\sum_A \theta_A^a \tilde G_A)~,~~~
U_b=\exp(\sum_A \theta_A^b \tilde G_A)~,~~~
U_{ba}=\exp(\sum_A \theta_A^{ba} \tilde G_A)~,...,~~~ \eqno(15a)$$
with $\theta_A^e=0$ and $U_e=1$.
Then Eq.~(4b) implies that $\Omega(b,a)$ must be unity when either $a$ or
$b$ is the identity, and thus the generator form for this operator is
$$\Omega(b,a)=\exp\left({1\over2} \sum_{BA}[\theta_B^b \theta_A^a \tilde I_{BA} 
+\sum_C\theta_B^b \theta_C^b \theta_A^a \tilde J^{(1)}_{(BC)A} 
+\sum_C \theta_B^b \theta_A^a \theta_C^a {\tilde J}^{(2)}_{B(AC)} 
+{\rm O}(\theta^4)]\right)~~~,\eqno(15b)$$
where the parentheses $(~)$ around a set of indices 
indicate that the tensor in 
question is symmetric in those indices, and where we use the     
tilde to indicate operators which are anti--self--adjoint.  The 
parameters $\theta_C^{ba}$ must be functions of the parameters $\theta_A^a$ 
and $\theta_B^b$, 
$$\theta_C^{ba}=\psi_C^{ba}(\{\theta_B^b \},\{ \theta_A^a \})
=\theta_C^b+ \theta_C^a + {1\over2} \sum_{BA} C_{BAC} \theta_B^b \theta_A^a
+{\rm O}(\theta^3)~~~,\eqno(15c)$$
where in making the Taylor expansion we have used the fact that $U_{be}=U_b$ 
and $U_{ea}=U_a$, which fixes the linear terms in the expansion and requires    
the quadratic term to be bilinear.  

We proceed now to derive a number of relations by combining the generator 
expansions of Eqs.~(15a--c) with the formulas of Sec.~I.  We begin by 
substituting Eqs.~(15a--c) into Eq.~(4b) using the Baker--Campbell--
Hausdorff formula,  
$$\exp X \exp Y=\exp (X+Y+{1\over2}[X,Y]+...)~~~,  \eqno(16a)$$
to combine exponents arising from the factors on the left and right.  
>From the left hand side of Eq.~(4b) we get
$$U_b U_a=\exp(\sum_B \theta^b_B \tilde G_B+\sum_A \theta^a_A \tilde G_A
+{1\over2} \sum_{BA} \theta_B^b \theta_A^a [\tilde G_B, \tilde G_A]
+{\rm O}(\theta^3))~~~,\eqno(16b)$$
while from the right hand side of Eq.~(4b) we get
$$U_{ba}\Omega(b,a)=\exp\left(\sum_C(\theta_C^b+\theta_C^a) \tilde G_C +
{1\over2} \sum_{CBA} C_{BAC} \theta_B^b \theta_A^a 
\tilde G_C + {1\over2} \sum_{BA} \theta_B^b \theta_A^a \tilde I_{BA}
+{\rm O}(\theta^3)\right)~~~.\eqno(16c)$$
Equating Eqs.~(16b) and (16c) thus gives the relations 
$$[\tilde G_B, \tilde G_A]=\sum_C C_{[BA]C} \tilde G_C + \tilde I_{[BA]}~~~ 
\eqno(17a)$$
and
$$0=\sum_C C_{(BA)C} \tilde G_C + \tilde I_{(BA)}~~~, 
\eqno(17b)$$
where the square brackets $[~]$ around a set of indices indicates that the 
tensor in question is antisymmetric in these indices.  We shall restrict 
ourselves henceforth to the case in which $C_{(BA)C}=0$, which by Eq.~(17b) 
implies that $\tilde I_{(BA)}=0$; making this assumption then implies that
$C_{BAC}=C_{[BA]C}$ and $\tilde I_{BA}=\tilde I_{[BA]}$.  In other words, 
we are assuming that the structure constants $C_{BAC}$ for a projective 
representation have the same antisymmetric form as holds for a vector 
representation.  
Changing the summation index $C$ to $D$ in 
Eq.~(17a), and then taking the commutator of Eq.~(17a) with $\tilde G_C$, 
we find
$$[\tilde G_C,[\tilde G_B, \tilde G_A]]
=\sum_D C_{[BA]D} [\tilde G_C, \tilde G_D]+ [\tilde G_C, \tilde I_{[BA]}]
~~~; \eqno(18a)$$
adding to this identity the two related identities obtained by cyclically 
permuting $A,B,C$, using the fact that the left hand side of the sum 
vanishes by the Jacobi identity for the commutator, and substituting Eq.~(17a)  
for the commutators appearing on the right hand side of the sum, we get 
the identity 
$$\eqalign{
&\sum_{DE}( C_{[BA]D} C_{[CD]E} + C_{[CB]D} C_{[AD]E} + C_{[AC]D} C_{[BD]E} )
\tilde G_E  \cr
+&\sum_D(C_{[BA]D} \tilde I_{[CD]} +C_{[CB]D} \tilde I_{[AD]}   
+C_{[AC]D} \tilde I_{[BD]} )  \cr
+&[\tilde G_C, \tilde I_{[BA]} ] + [\tilde G_A, \tilde I_{[CB]} ]
+[\tilde G_B, \tilde I_{[AC]} ]=0~~~.
}\eqno(18b)$$

We next substitute Eqs.~(15a--c) into the associativity condition of 
Eq.~(8), now keeping cubic terms in the exponent of the form 
$\theta_A^a \theta_B^b \theta_C^c$, but dropping cubic 
terms, such as $\theta_A^a \theta_B^a \theta_C^c$, that 
do not contain all three of the upper indices $a,b,c$.  
For the first factor on the 
left hand side of Eq.~(8), we find from Eqs.~(15b) and (15c) that
$$\eqalign{
\Omega(c,ba)=&\exp\left({1\over2} \sum_{BA}(\theta_B^c \theta_A^{ba} \tilde I_{[BA]}
+\sum_C \theta_B^c \theta_A^{ba} \theta_C^{ba} {\tilde J}^{(2)}_{B(AC)}) \right) \cr
=&\exp\left({1\over2} \sum_{BA}[\theta_B^c 
(\theta_A^b+\theta_A^a+{1\over2}\sum_{DE} C_{[DE]A} \theta_D^b \theta_E^a) 
\tilde I_{[BA]}
+2 \sum_C \theta_B^c \theta_A^b
\theta_C^a {\tilde J}^{(2)}_{B(AC)} ]\right)~~~, \cr
}\eqno(19a)$$
while for the second factor on the left hand side of Eq.~(8) we have  
$$\Omega(b,a)=\exp({1\over2} \sum_{BA}\theta_B^b \theta_A^a \tilde I_{[BA]})
~~~.\eqno(19b)$$
Since the exponents in Eqs.~(19a, b) both begin at order $\theta^2$, through 
order $\theta^3$ we can simply add exponents to get the product on the left 
hand side of Eq.~(8).  Proceeding similarly for the first factor on the 
right hand side of Eq.~(8), we get
$$\eqalign{
\Omega(cb,a)=&\exp\left({1\over2} \sum_{BA}(\theta_B^{cb} \theta_A^a \tilde I_{[BA]}
+\sum_C\theta_B^{cb} \theta_C^{cb} \theta_A^a \tilde J^{(1)}_{(BC)A} )\right) \cr
=&\exp\left({1\over2} \sum_{BA}[
(\theta_B^c+\theta_B^b +{1\over2} \sum_{DE} C_{[DE]B}\theta_D^c \theta_E^b) 
\theta_A^a \tilde I_{[BA]}
+2 \sum_C\theta_B^c \theta_C^b \theta_A^a \tilde J^{(1)}_{(BC)A} ]\right)~~~, \cr
}\eqno(20a)$$
while for the second factor on the right hand side of Eq.~(8), use of  
the Baker--Campbell--Hausdorff formula gives 
$$\eqalign{
U_a^{-1} \Omega(c,b) U_a=&\exp(-\sum_A \theta_A^a \tilde G_A)
\exp({1\over2} \sum_{CB}\theta_C^c \theta_B^b \tilde I_{[CB]} )
\exp(\sum_A \theta_A^a \tilde G_A)\cr
=&\exp({1\over2} \sum_{CB}\theta_C^c \theta_B^b \tilde I_{[CB]}
-{1\over2} \sum_A\sum_{CB} \theta_A^a \theta_C^c \theta_B^b
[\tilde G_A, \tilde I_{[CB]} ] ) ~~~. \cr
}\eqno(20b)$$
Since the exponents in Eqs.~(20a, b) begin at order $\theta^2$, it again 
suffices to simply add the exponents to form the product appearing on the 
right hand side of Eq.~(8).  Thus, to the requisite order, the content of 
Eq.~(8) is obtained by equating the sum of the exponents in Eqs.~(19a, b) to 
the corresponding sum of exponents in Eqs.~(20a, b).  The quadratic terms 
in $\theta$ are immediately seen to be identical on left and right, while 
the cubic term proportional to $\theta_A^a \theta_B^b \theta_C^c$ gives 
(after some relabeling of dummy summation indices) the nontrivial identity
$$\tilde J_{C(BA)}^{(2)} +{1\over 4} \sum_D C_{[BA]D} \tilde I_{[CD]}
=\tilde J_{(CB)A}^{(1)} +{1\over 4} \sum_D C_{[CB]D} \tilde I_{[DA]}
-{1 \over 2} [\tilde G_A, \tilde I_{[CB]} ] ~~~.\eqno(21)$$
On totally antisymmetrizing with respect to the indices $A,B,C$, the 
terms in Eq.~(21) involving $\tilde J^{(1,2)}$ drop out, and we are left with 
the identity
$$\eqalign{
&\sum_D(C_{[BA]D} \tilde I_{[CD]} +C_{[CB]D} \tilde I_{[AD]}   
+C_{[AC]D} \tilde I_{[BD]} )  \cr
+&[\tilde G_C, \tilde I_{[BA]} ] + [\tilde G_A, \tilde I_{[CB]} ]
+[\tilde G_B, \tilde I_{[AC]} ]=0~~~.\cr
}\eqno(22a)$$
In other words, associativity implies that the sum of the 
second and third lines of 
Eq.~(18b) vanishes separately; hence the first line of Eq.~(18b) must also 
vanish, and since the generators $\tilde G_E$ are linearly independent this 
gives the Jacobi identity for the structure constants, 
$$\eqalign{
&\sum_{DE}( C_{[BA]D} C_{[CD]E} + C_{[CB]D} C_{[AD]E} + C_{[AC]D} C_{[BD]E} )
=0~~~. \cr
}\eqno(22b)$$
In the complex case, in which $\Omega(a,b) =\omega(a,b)1$ is a $c$--number, 
the tensor $\tilde I_{[AB]}$ is a $c$--number ``central charge'' and the 
commutator terms in Eqs.~(18b) and (22a) vanish identically.  Therefore,  
in the complex case,  
Eq.~(18b) implies both Eq.~(22b) and the identity 
$$\sum_D(C_{[BA]D} \tilde I_{[CD]} +C_{[CB]D} \tilde I_{[AD]}   
+C_{[AC]D} \tilde I_{[BD]} )=0~~~  {\rm complex~case},  \eqno(23)$$
and so one obtains the entire 
content of the associativity condition 
from the simpler analysis leading to Eq.~(18b), without having to 
perform the third order expansion needed to get Eq.~(22a).

\leftline{\bf III.~~GENERAL, MULTI--CENTRAL, AND CENTRAL QUATERNIONIC} 
\leftline{\bf ~~~~~~PROJECTIVE REPRESENTATIONS}

The analysis of Sec.~II applies to the general case (apart from the 
restriction $C_{(BA)C}=0$) of a quaternionic 
projective representation; in order to obtain more detailed results 
it is necessary to introduce further structural assumptions.  In 
Ref.~4 two special classes of quaternionic projective representations 
are defined.  A quaternionic projective representation is defined to 
be {\it multi--central} if
$$[\Omega(b,a), U_a]=[\Omega(b,a), U_b]=0~,~~~{\rm all}~a,b~~~,\eqno(24a)$$
while it is defined to be {\it central} if 
$$[\Omega(b,a), U_c]=0~,~~~{\rm all}~a,b,c~~~.\eqno(24b)$$
Expressed in terms of the generators introduced in Eqs.~(15a, b), the 
multi--centrality condition takes the form 
$$\sum_{ABC} \theta_A^a \theta_B^b \theta_C^a [\tilde G_C, \tilde I_{[BA]} ] 
= \sum_{ABC} \theta_A^a \theta_B^b \theta_C^b [\tilde G_C, \tilde I_{[BA]} ]
=0~,~~~{\rm all}~a,b~~~,\eqno(25a)$$  
while the centrality condition becomes
$$\sum_{ABC} \theta_A^a \theta_B^b \theta_C^c [\tilde G_C, \tilde I_{[BA]} ]
=0~,~~~{\rm all}~a,b,c~~~.\eqno(25b)$$  
Making the definition 
$$\Delta_{[AB]C}=[\tilde G_C, \tilde I_{[BA]} ]~~~,\eqno(25c)$$
we see from Eq.~(25a) that multi--centrality 
requires that $\Delta_{[AB]C}$ be 
antisymmetric in $A,C$ and in $B,C$ as well as in $A,B$; thus in   
the multi--central case $\Delta$ is totally antisymmetric, which we will 
indicate by writing it as $\Delta_{[ABC]}$.   
>From Eq.~(25b), we see that centrality requires that  
$\Delta_{[AB]C}$ must vanish.  

Using the generator formulation, we proceed now 
to discuss successively the general, multi--central, and central cases in the 
light of the associativity analysis of Sec.~II. \hfill\break
\parindent=0pt
(1)  {\it The general case.}  An example given in Eqs.~(13.54g) and (14.23a)  
of Ref.~4 shows that one can have a quaternionic projective 
representation which is neither central nor multi--central.  The example 
is constructed from $n$ independent fermion creation and annihilation 
operators $b^{\dagger}_{\ell},~b_{\ell}, ~~\ell=1,...,n$, 
which commute with a left algebra quaternion basis $E_0=1,E_1=I,E_2=J,E_3=K$.
Consider the set of three generators $\tilde G_A$ defined by 
$$\tilde G_A=-{1 \over 2} E_A N~,~~~A=1,2,3~~~,\eqno(26a)$$
with $N$ the number operator 
$$N=\sum_{\ell=1}^n  b_{\ell}^{\dagger} b_{\ell}~~~.\eqno(26b)$$
The commutator algebra of these generators has the form of a projective 
representation of $SU(2)$, 
$$\eqalign{
[\tilde G_B, \tilde G_A]=& -\sum_{C=1}^3 \epsilon_{[BAC]} \tilde G_C 
+\tilde I_{[BA]}~,~~~\cr
\tilde I_{[BA]}=&\sum_{C=1}^3 \epsilon_{[BAC]} {1 \over 2} E_C N (N-1)~~~, \cr
}\eqno(26c)$$
with $\epsilon$ the usual three index antisymmetric tensor.  
A simple calculation now shows that 
$$[\tilde G_A, \tilde I_{[BC]}]=-N(N-1)(\delta_{AB}\tilde G_C - \delta_{AC}
\tilde G_B)~~~,\eqno(27a)$$
which is not antisymmetric in either the index pair $A,C$ or the pair 
$A,B$, and so the multi--centrality condition is not satisfied.   
Another simple calculation shows that 
$$\sum_D (\epsilon_{[BAD]} \tilde I_{[CD]} + \epsilon_{[CBD]} \tilde I_{[AD]}
+\epsilon_{[ACD]} \tilde I_{[BD]})=0~~~,\eqno(27b)$$
by virtue of the Jacobi identity for the structure constant $\epsilon$, 
and also 
$$[\tilde G_C , \tilde I_{[BA]} ] + [\tilde G_A , \tilde I_{[CB]} ]
+[\tilde G_B , \tilde I_{[AC]} ]  =0~~~.\eqno(27c)$$
Hence the associativity condition of Eq.~(22a) is satisfied, with the first 
and second lines each vanishing separately.  \hfill\break
\parindent=0pt
(2)  {\it The multi--central case.}  Let us now consider the multi--central 
case, in which $\Delta_{[AB]C}$ defined in Eq.~(25c) is totally 
antisymmetric in 
$A,B,C$, as indicated by the notation $\Delta_{[ABC]}$.  
The associativity condition of Eq.~(22a) then simplifies to 
$$\sum_D(C_{[BA]D} \tilde I_{[CD]} +C_{[CB]D} \tilde I_{[AD]}   
+C_{[AC]D} \tilde I_{[BD]} )  +3 \Delta_{[ABC]}=0~~~.
\eqno(28a)$$
A further equation involving $\Delta$ is obtained from the Jacobi identity 
$$[\tilde G_D, [\tilde G_C, \tilde I_{[BA]}]]   
- [\tilde G_C, [\tilde G_D, \tilde I_{[BA]}]]   
=[\tilde I_{[BA]},[\tilde G_C, \tilde G_D]]~~~,\eqno(28b)$$
which on substituting Eqs.~(17a) and (25c) becomes
$$[\tilde G_D,\Delta_{[AB]C}]-[\tilde G_C,\Delta_{[AB]D}]
=-\sum_E C_{[CD]E} \Delta_{[AB]E} + [\tilde I_{[BA]}, \tilde I_{[CD]}]~~~,
\eqno(28c)$$
an equation which holds even in the general case in which $\Delta$ is 
not totally antisymmetric.   Specializing Eq.~(28c) to the multi--central 
case and contracting it with $\delta_{AC} \delta_{BD}$, the left hand side 
vanishes because of the antisymmetry of $\Delta$, while the commutator 
term on the right hand side becomes $\sum_{AB} [\tilde I_{[BA]}, 
\tilde I_{[AB]}]=0$, leaving the identity (after relabeling the dummy index 
$E$ as $C$)
$$\sum_{ABC} C_{[AB]C} \Delta_{[ABC]}=0~~~.\eqno(29)$$
Thus in order for a multi--central projective representation to exist which 
has $\Delta \not =0$ and so is not also central, there must be a three 
index antisymmetric tensor $\Delta_{[ABC]}$ 
which vanishes when all three indices are 
contracted with the structure constant $C_{[AB]C}$.  This condition is not 
easy to satisfy and so we pose the question, which we have not been 
able to answer:  Can one construct 
an example of a multi--central quaternionic 
projective representation which is not central, or can one prove 
(in general, or with a restriction, e.g., to simple or semi--simple groups)  
that a multi--central quaternionic projective 
representation must always be central?  The 
application of multi--centrality in Ref.~4 sheds no light on this 
issue; multi--centrality was used there (e.g. in Sec. 12.3) 
to show that quaternionic 
Poincar\'e group projective representations outside the zero energy sector
can always be transformed to complex Poincar\'e group projective 
representations, which in the sector continuously connected to the identity 
are known$^8$ to be transformable to vector representations. 
\hfill\break
\parindent=0pt
(3)  {\it The central case.}  Let us finally consider the central 
case in which $\Delta=0$, which by Eqs.~(25c) and (28c) implies that 
$\tilde I_{[BA]}$ commutes with both $\tilde G_C$ and $\tilde I_{[CD]}$ 
for arbitrary values of the indices.  Thus $\tilde I_{[BA]}$ behaves as a   
central charge, justifying the name ``central'' for  this case.  The various 
results obtained in  Bargmann$^6$ can be immediately generalized to 
the quaternionic central case; for example, the analysis of Ref.~6 
can be easily extended to show that the central  
charges associated with a 
quaternionic central projective representation of a semi--simple Lie group 
can always be removed by redefinition of the generators; and again, the 
nontrivial illustration$^6$ of 
a complex projective representation, constructed in terms of the phase space 
translation generators in nonrelativistic quantum mechanics, can be 
embedded$^4$ 
in quaternionic quantum mechanics as a central projective 
representation. 
\bigskip
\leftline{\bf ACKNOWLEDGMENTS}
This work was supported in part by the Department of Energy under
Grant \#DE--FG02--90ER40542.  I wish to acknowledge the hospitality of the
Aspen Center for Physics, and of the Department of Applied Mathematics and
Theoretical Physics and Clare Hall at Cambridge University, where parts of
this work were done, and wish to thank E. Witten for helpful conversations.  
\hfill \break
\bigskip
\leftline{\bf REFERENCES}
\bigskip
\noindent
1.~~G. Emch and C. Piron, J. Math. Phys. {\bf 4}, 469 (1963). 
\bigskip
\noindent
2.~~U. Uhlhorn, Arkiv Phys. {\bf 23}, 307 (1963). 
\bigskip
\noindent
3.~~V. Bargmann, J. Math. Phys. {\bf 5}, 862 (1964).
\bigskip
\noindent
4.~~S.L. Adler  {\it Quaternionic Quantum Mechanics and Quantum Fields} 
(Oxford University Press, New York, 1995); see especially Sec.~4.3.
\bigskip 
\noindent
5.~~G. Emch, Helv. Phys. Acta {\bf 36}, 739, 770 (1963). 
\bigskip
\noindent
6.~~V.Bargmann,  Ann. Math. {\bf 59}, 1 (1954).
\bigskip
\noindent
7.~~S. Weinberg {\it The Quantum Theory of Fields, Vol. 1}  (Cambridge 
University Press, Cambridge, 1995).
\bigskip
\noindent
8.~~E. P. Wigner, Ann. Math. {\bf 40}, 149 (1939).  
\bigskip
\vfill
\eject
\bye